# Soliton control in chirped photonic lattices

Yaroslav V. Kartashov*

*ICFO-Institut de Ciencies Fotoniques, 08034 Barcelona, Spain*

Victor A. Vysloukh

*Departamento de Fisica y Matematicas, Universidad de las Americas – Puebla, Santa Catarina Martir, 72820 Puebla, Mexico*

Lluis Torner

*ICFO-Institut de Ciencies Fotoniques, and Department of Signal Theory and Communications, Universitat Politecnica de Catalunya, 08034 Barcelona, Spain*

We study optical solitons in chirped periodic optical lattices whose amplitude or frequency changes in the transverse direction. We discover that soliton propagation in such lattices can be accompanied by the progressive self-bending of the soliton trajectory, and we show that the soliton bending rate and output position can be controlled by varying the lattice depth, as well as the chirp amplitude and frequency modulation rate. This effect has potential applications for controllable soliton steering and routing.

Micro-structured optical devices (e.g., chirped fiber Bragg gratings and mirrors, arrayed waveguide gratings) are effective tools for the generation and processing of ultra-short optical pulses (see, e.g., Refs 1-2 and papers quoted therein). Such devices found applications in many settings, including wavelength-stabilized lasers, Raman amplifiers, phase conjugators, passive optical networks, or dispersion compensators.[3,4] For example, properly designed chirped gratings can be used to compensate the group velocity mismatch experienced by waves of different frequencies. The close analogy between dispersion and diffraction effects suggests that ideas born in ultrafast optics might be transferred from a time-domain to a space-domain in order to control diffraction of laser beams, in particular, to manage dynamics and properties of spatial solitons. This can be accomplished in periodic waveguide arrays with controllable refractive index modulation depth and waveguide separation, a feature made possible by the advent of optically-induced lattices in photorefractive media.[5-13] The diffractive properties of such lattices can be tuned to a great extent by varying intensity of light waves employed to induce them, thus opening a promising avenue for spatial soliton control concepts like radiative switching and parametric steering.[14-18]

The recent, previous studies of this possibility addressed periodic (unchirped) lattices with the constant refractive index modulation depth. In this paper we uncover additional possibilities for soliton control that are accessible with periodic lattices whose amplitude or frequency is modulated linearly in the transverse direction. We show that soliton propagation in such lattice is accompanied by progressive light bending in the direction of growth of the lattice amplitude or decrease of its local frequency. The soliton bending and consequently the output soliton position can be controlled by varying the lattice depth, its spatial frequency, and the amplitude or frequency modulation rate, a feature with direct potential applications for controllable soliton steering and switching.

We consider propagation of optical radiation along the $\xi$ axis in cubic nonlinear medium with modulation of linear refractive index along transverse $\eta$ axis, described by the nonlinear Schrödinger equation for dimensionless complex field amplitude $q$:

$$i\frac{\partial q}{\partial \xi} = -\frac{1}{2}\frac{\partial^2 q}{\partial \eta^2} - q|q|^2 - pR(\eta)q. \qquad (1)$$

Here the longitudinal $\xi$ and transverse $\eta$ coordinates are scaled to the diffraction length and input beam width, respectively. The parameter $p$ is proportional to the depth of refractive index modulation, while the function $R(\eta)$ stands for the transverse profile of the refractive index. We consider optical lattices with linear amplitude modulation (AM lattices), whose transverse profile is described by the function $R(\eta) = (1+\alpha\eta)\cos(\Omega_\eta \eta)$, where $\alpha < 1$ is the modulation rate and $\Omega_\eta$ is the modulation frequency, and lattices with linear frequency modulation (FM lattices) described by $R(\eta) = \cos[\Omega_\eta \eta(1+\alpha\eta)]$. For convenience, we use the same notation for the dimensionless modulation rate $\alpha$ for both, AM and FM lattices.

Besides the possibility of direct technological fabrication of refractive index profiles with parameters that vary linearly in transverse direction, it should be mentioned that at $\alpha\eta \ll 1$ AM-lattices are identical to lattices with harmonic refractive index modulation $R(\eta) = [1 + \sin(\alpha\eta)]\cos(\Omega_\eta \eta)$ that can be induced optically, e.g. in photorefractive media, with several interfering plane waves. The technique of optical lattice induction provides an important possibility to tune lattice parameters, including depth and frequency of the refractive index modulation. We assume that the depth of the refractive index modulation is small compared to the unperturbed index and is of the order of the nonlinear correction to the refractive index due to the Kerr effect, so that lattices addressed here are relatively shallow and thus, in general, the tight-binding approximation can not be applied for their analysis. Eq. (1) admits several conserved quantities, including the power, or energy flow, $U = \int_{-\infty}^{\infty} |q|^2 \, d\eta$.

First we address the properties of stationary soliton solutions supported by linearly modulated lattices. We search for soliton solutions in the form $q(\eta,\xi) = w(\eta)\exp(ib\xi)$, where $w(\eta)$ is the real function and $b$ is the propagation constant. We find profiles of solitons located in the vicinity of the point $\eta = 0$ numerically from Eq. (1) with relaxation method. To analyze the dynamic stability of the obtained soliton families we searched for the perturbed solutions of Eq. (1) in the form $q(\eta,\xi) = [w(\eta) + u(\eta,\xi) + iv(\eta,\xi)]\exp(ib\xi)$, where perturbation components $u,v$ can grow upon propagation with a complex growth rate $\delta$. Linearization of Eq. (1) around a stationary solution $w(\eta)$ yields a linear eigenvalue problem that we solved numerically. Here we are interested only in the simplest ground-state soliton solutions.

The salient properties of solitons supported by AM lattices are summarized in Fig. 1. The energy flow $U$ is a nonmonotonic function of the propagation constant $b$ (Fig. 1(a)). There exist a lower propagation constant cutoff $b_{co}$ for soliton existence. Physically this cutoff arises due to the competition between the harmonic refractive index modulation and the linear increase of lattice amplitude at $\eta \to \infty$. Actually, in the absence of the harmonic modulation ($\Omega_\eta = 0$) a soliton launched into a nonlinear medium with $\alpha \neq 0$ would travel towards the positive direction of the $\eta$ axis (in the direction of increase of the refractive index), while harmonic modulation introduces a potential barrier that prevents soliton from traveling and makes possible the very existence of stationary soliton solutions in AM lattices. Such potential barrier arises because of the nonlinearity of the medium response and does not exist in linear lattices, where progressive coupling of light into regions with higher refractive index is unavoidable. Since the height of the potential barrier in the nonlinear lattice depends on the energy flow and on the width of the beam, low-energy stationary solitons are not supported by lattices; hence, the existence of a lower cutoff. The cutoff monotonically increases with increase of lattice depth (Fig. 1(b)) and linear amplitude modulation rate (Fig. 1(c)). At high energy flow levels soliton profiles almost symmetric, while close to the cutoff for existence they are distorted toward regions with higher refractive index (Fig. 1(d)).

The outcome (numerical) of the linear stability analysis revealed that ground-state solitons are stable almost in the entire domain of their existence except for the narrow region near the cutoff where $dU/db \leq 0$. This can be viewed as a confirmation of the applicability of Vakhitov-Kolokolov stability criterion for ground-state soliton solutions of Eq. (1). For a fixed set of parameters $\alpha$, $p$, and $\Omega_\eta$ the properties of solitons depend also on the $\eta$-location of its intensity maximum. In particular, the propagation constant cutoff $b_{co}$ is higher for solitons whose centers are shifted in the positive direction of the $\eta$-axis. Nevertheless, we found that the qualitative character of dependencies $U(b)$, $b_{co}(p)$, and $b_{co}(\alpha)$ are not affected by the position of the soliton center as long as $\eta > -1/\alpha$.

The most interesting situation arises when the energy flow of the input beam is not sufficient for formation of stationary soliton for a given width and lattice parameters. In this case the periodic modulation cannot prevent beam from bending toward region with higher refractive index, but in contrast to medium with linearly growing refractive index ($\Omega_\eta = 0$), the bending rate of beam inside the lattice depends on its frequency and depth (Fig. 2). This opens the possibility for control of output beam center position by tuning the parameters of the lattice, a goal that can be easily realized in the case of optically-induced lattices.

Next we study the propagation of beams traveling across the lattice. We solve Eq. (1) with the input conditions $q|_{\xi=0} = \text{sech}(\eta)\exp(i\nu\eta)$, where $\nu$ is the incident angle. This choice of the input conditions is justified since it corresponds to exact soliton solution in the homogeneous case and enables to minimize the radiative losses at the

initial stage of propagation. Suppression of radiative losses is directly connected with enhanced mobility of such broad input beams that cover several lattice periods. Note that the trajectory of broad traveling beam is almost parabolic, beam does not broaden upon propagation since nonlinearity compensates diffraction (in this sense, such traveling beam are called soliton), and radiation that unavoidably arises when soliton crosses lattice channels is weak if local propagation angle with respect to $\xi$ axis is far from Bragg angle. It should be pointed out that the propagation trajectories of narrow input beams, whose width is comparable with the lattice period, may depart considerably from parabolic ones. We did not observe periodic Bloch oscillations that are known to occur in discrete waveguide arrays with linearly increasing refractive index in neighboring waveguides.[19,20] In contrast, soliton beam was found to be destroyed when its local propagation angle approached the Bragg one. This is because of the peculiar difference between the structure of AM lattice addressed here and waveguide arrays of Refs [19,20], whose profiles incorporate the sum of linear and periodic refractive index modulation. An increase in lattice amplitude modulation rate leads to monotonic increase of soliton bending rate (Fig. 2(a)). The shift of integral soliton center, defined as:

$$\delta\eta = \frac{1}{U}\int_{-\infty}^{\infty} \eta|q|^2\,d\eta \qquad (2)$$

at $\xi = 32$ is a nonmonotonic function of modulation frequency (Fig. 2(b)). At $\Omega_\eta < 6$ the input beam forms stationary immobile soliton, while at $\Omega_\eta > 6$ it starts to travel across the lattice and bending rate reaches its maximal value for $\Omega_\eta \approx 7.2$. The rate of bending monotonically decreases as $\Omega_\eta \to \infty$ that is consistent with the fact that broad solitons are almost unaffected by high-frequency refractive index modulation. The soliton center shift increases as $p^2$ (Fig. 2(c)) that can be viewed as another manifestation of controlled bending afforded by AM lattices. Finally, the possible propagation trajectories can be enriched by launching soliton at nonzero angle $\nu$ with respect to the lattice (Fig. 2(d)). Thus, for high enough negative $\nu$ soliton can penetrate the area $\eta < -1/\alpha$ where it will experience bending in negative direction of $\eta$ axis. As one can see from Figs. 2(a) and 2(d), the soliton center shift can be quite considerable (of the order of several soliton width) already at propagation distance $\xi \sim 16$. Notice that in photorefractive crystals used for experimental generation of optical lattices,[6,7,10-12] such as SBN biased with a static electric field of some $E \sim 10^5$ V/cm, for a beam with width $r_0 \sim 10$ $\mu$m at the wavelength $\lambda = 0.63$ $\mu$m, the distance $\xi \sim 16$ would correspond to actual propagation length about $32$ mm. Therefore the effect of controlled soliton bending should be observable in such crystals.

The main properties of stationary solitons supported by FM lattices are summarized in Fig. 3. Dependencies $U(b)$ and $b_{co}(p)$ for such solitons are very similar to that for solitons supported by AM lattice. At high energy flows solitons are almost symmetric, while close to cutoff they become distorted in the direction of decrease of local lattice frequency $\Omega_\eta(1+2\alpha\eta)$ (Fig. 3(a)). As in the case of AM lattices the shift of soliton center location along $\eta$-axis does not qualitatively affect soliton properties. Despite the fact that optical-induction of FM lattices is not obvious, they offer a number of unique opportunities for soliton steering, since widths of guiding channels in such lattices change in transverse direction, so that soliton mobility changes across the lattice. A soliton $q|_{\xi=0} = \mathrm{sech}(\eta)$ launched into a FM lattice experiences the refractive index distribution $\delta n \sim p\,\mathrm{sech}[\pi\Omega_\eta(1+2\alpha\eta)/2]$ averaged over fast oscillations and, therefore, experiences attraction to the zero-frequency point $\eta = -1/2\alpha$ (Fig. 3(c)). Intuitively, such soliton starts to travel across the lattice in the direction of decrease of local frequency and finally can be trapped in the guiding lattice channel whose width somehow matches the soliton width. The distance where trapping occurs rapidly decreases with increase of lattice depth, while the position of output channel remains unchanged (Fig. 3(c)). Note that soliton never passes the point $\eta = -1/2\alpha$, where local frequency of the lattice goes to zero. Finally, by changing the incident angle $\nu$ it is possible to address the output guiding channel at different propagation distances (Fig. 3(d)). The potential of the scheme for spatial soliton manipulation is clearly apparent.

We thus conclude stressing that we have exposed that the amplitude and frequency modulation of transversally-chirped photonic lattices offers important new opportunities for the diffraction control and soliton steering. The key feature uncovered here is the possibility to control the soliton mobility across the lattice by proper selection of the chirp modulation rate and the overall lattice parameters.

*On leave from Physics Department of M. V. Lomonosov Moscow State University, Russia. This work has been partially supported by the Government of Spain through BFM2002-2861 and by the Ramon-y-Cajal Program.

# Figure captions

Figure 1. (a) Energy flow versus propagation constant at $p=4$. (b) Cutoff versus lattice depth at $\alpha=0.2$. Inset shows lattice profile. (c) Cutoff versus lattice amplitude modulation rate at $p=2$. (d) Profiles of solitons with different energy flows at $p=4$, $\alpha=0.2$. Gray regions in (d) correspond to $R(\eta)\leq 0$ and white regions correspond to $R(\eta)>0$. Modulation frequency $\Omega_\eta=8$.

Figure 2. (a) Propagation dynamics of solitons in lattices with different amplitude modulation rates at $p=2$, $\Omega_\eta=8$. (b) Soliton center shift at distance $\xi=32$ versus modulation frequency at $p=2$, $\alpha=0.1$. (c) Soliton center shift at distance $\xi=32$ versus lattice depth at $\Omega_\eta=8$, $\alpha=0.1$. (d) Propagation dynamics of solitons with different input angles at $p=1$, $\Omega_\eta=8$, $\alpha=0.3$.

Figure 3. (a) Profiles of solitons with different energy flows at $p=4$, $\alpha=0.05$. Gray regions in (a) correspond to $R(\eta)\leq 0$ and white regions correspond to $R(\eta)>0$. (b) Cutoff versus lattice depth at $\alpha=0.05$. Inset shows lattice profile. (c) Propagation dynamics of solitons in lattices with different depths at $\alpha=0.02$. (d) Propagation dynamics of solitons with different input angles at $p=2$, $\alpha=0.02$. Modulation frequency $\Omega_\eta=8$.

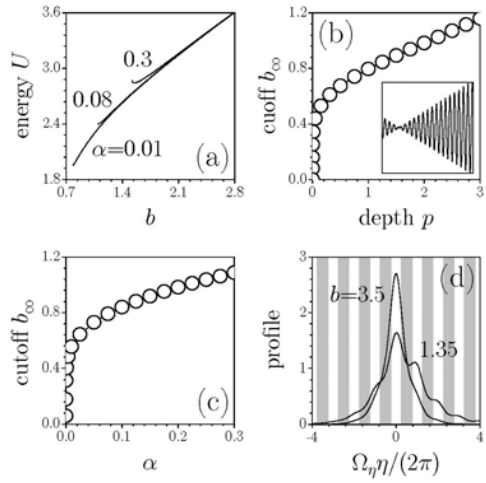

Figure 1. (a) Energy flow versus propagation constant at $p=4$. (b) Cutoff versus lattice depth at $\alpha=0.2$. Inset shows lattice profile. (c) Cutoff versus lattice amplitude modulation rate at $p=2$. (d) Profiles of solitons with different energy flows at $p=4$, $\alpha=0.2$. Gray regions in (d) correspond to $R(\eta)\leq 0$ and white regions correspond to $R(\eta)>0$. Modulation frequency $\Omega_\eta=8$.

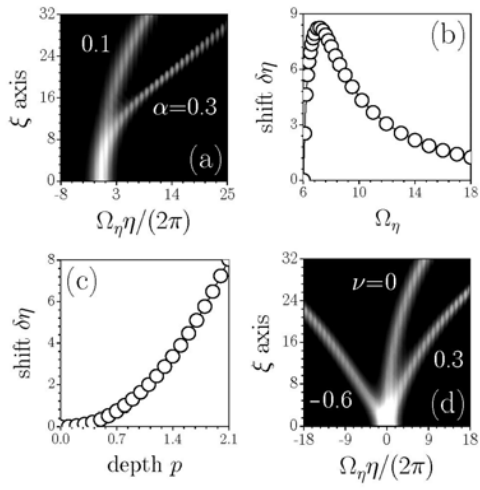

Figure 2. (a) Propagation dynamics of solitons in lattices with different amplitude modulation rates at $p=2$, $\Omega_\eta = 8$. (b) Soliton center shift at distance $\xi = 32$ versus modulation frequency at $p=2$, $\alpha = 0.1$. (c) Soliton center shift at distance $\xi = 32$ versus lattice depth at $\Omega_\eta = 8$, $\alpha = 0.1$. (d) Propagation dynamics of solitons with different input angles at $p=1$, $\Omega_\eta = 8$, $\alpha = 0.3$.

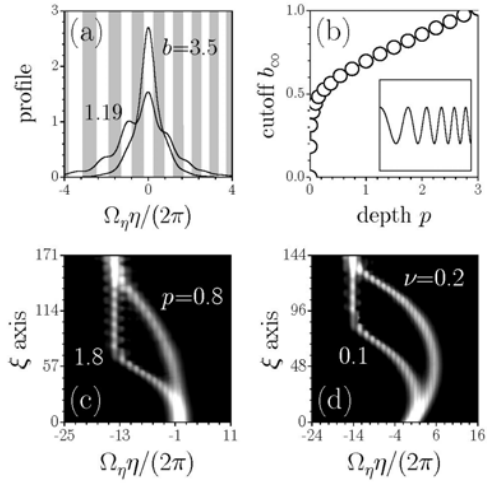

Figure 3. (a) Profiles of solitons with different energy flows at $p=4$, $\alpha=0.05$. Gray regions in (a) correspond to $R(\eta)\leq 0$ and white regions correspond to $R(\eta)>0$. (b) Cutoff versus lattice depth at $\alpha=0.05$. Inset shows lattice profile. (c) Propagation dynamics of solitons in lattices with different depths at $\alpha=0.02$. (d) Propagation dynamics of solitons with different input angles at $p=2$, $\alpha=0.02$. Modulation frequency $\Omega_\eta=8$.